\documentclass[referee]{aa} 
\usepackage{graphicx}
\usepackage{txfonts}
\begin{document}
   \title{Intensity oscillations in coronal XBPs
from Hinode/XRT observations}

   \author{R. Kariyappa and B. A. Varghese}

   \institute{Indian Institute of Astrophysics, Bangalore 560 034, India\\
              \email{rkari@iiap.res.in}}
\offprints{R. Kariyappa}

   \date{Received ........................ / Accepted ....................... }

 
  \abstract
{}  
   {Our aim is to investigate the intensity oscillations in 
coronal X-ray Bright Points (XBPs).}
   {We analysed a 7-hours long time sequence of the
soft X-ray images obtained on April 14, 2007 with 2-min cadence using X-Ray Telescope
(XRT) on-board the Hinode mission.  We use SSW in IDL to derive the time series of 14 XBPs and 2
background regions. For the first time, we have tried to use power spectrum analysis on XBPs data to
determine the periods of intensity oscillations.}
{The power spectra of XBPs show several significant peaks at different
frequencies corresponding to a wide variety of time scales which range from a few minutes to hours.  The light curves 
of all the XBPs give the impression that the XBPs can be grouped into three classes
depending on emission levels: (i) weak XBPs; (ii) bright XBPs; and (iii) 
very strong XBPs.  The periods of intensity oscillation are consistent in all the XBPs and are independent of 
their brightness level, suggesting that the heating mechanisms in all the three groups of 
XBPs are similar.  The different classes of XBPs may be
related to the different strengths of the magnetic field with which they have been associated.}
   {}


\titlerunning {Intensity Oscillations in XBPs}
\authorrunning {R. Kariyappa and B.A. Varghese}
   \maketitle
%

\section{Introduction}

Solar coronal X-ray bright points (XBPs) have been an enigma since their discovery in late 1960's 
(Vaiana et al. 1970).
XBPs have subsequently been studied in great detail using Skylab and  Yohkoh X-ray images 
(Golub et al. 1974; Harvey 1996; Nakakubo \& Hara 1999; Longcope et al. 2001; Hara \& Nakakubo 2003).  
Their correspondence with small bipolar magnetic regions was
discovered by combining ground-based magnetic field measurements with simultaneous space-born
X-ray imaging observations (Krieger et al. 1971; Golub et al. 1977).
The number of XBPs (daily) found on the Sun varies from several
hundreds up to a few thousands (Golub et al. 1974).  Zhang et al. (2001) found a density of
800 XBPs for the entire solar surface at any given time.  It is known that the observed XBP number
is anti-correlated with the solar cycle, but this is an observational bias and the number
density of XBPs is nearly independent of the 11-yr solar activity cycle (Nakakubo \& Hara 1999;
Sattarov et al. 2002; Hara \& Nakakubo 2003).
Golub et al. (1974) found that the diameters of the
XBPs are around 10-20 arc sec and their life times range from 2 hours to 2 days (Zhang et al. 2001).  Studies 
have indicated the temperatures to be fairly low, $T= 2 \times 10^6$ K, and
the electron densities
{$n_e$ }= 5 $\times 10^{9} cm^{-3}$ (Golub \& Pasachoff 1997), although cooler XBPs exist (Habbal 1990).
XBPs are also useful as tracers of coronal rotation (Kariyappa 2008) and contribute to the Solar X-ray irradiance
variability (DeLuca and Saar 2008; Kariyappa \& DeLuca 2008).
Assuming that almost all XBPs represent new magnetic flux emerging at the solar surface, their overall
contribution to the solar magnetic flux would exceed that of the active regions (Golub \&
Pasachoff 1997).  Since a statistical interaction of the magnetic field is associated
with the production of XBPs, the variation of the XBP number on the Sun will be a measure
of the magnetic activity of its origin.

Bright points are also observed in the chromosphere using high resolution CaII H and K spectroheliograms and
filtergrams.  Extensive studies have been conducted to determine their dynamical evolution, the contribution to
chromospheric oscillations and heating, and to UV irradiance
variability (e.g. Liu 1974; Cram and Dam\'e 1983; Kariyappa et al. 1994; Kariyappa 1994 \& 1996; Kariyappa \&
Pap 1996; Kariyappa 1999; Kariyappa et al. 2005).  The oscillations of the
bright points at the higher chromosphere have been investigated using SOHO/SUMER Lyman series
observations (Curdt \& Heinzel 1998; Kariyappa et al. 2001).  It is known from these studies that the chromospheric
bright points are associated with 3-min period in their 
intensity variations using power spectra analysis.

Oscillations have been investigated in XBPs using Yohkoh/SXT (Strong et al. 1992) and found intensity variations
on time scales of a few minutes to hours. In 
2002, we (Kariyappa and Watanabe) made an
attempt to investigate the intensity oscillations at the sites of XBPs using Yohkoh/SXT observations.  We
found that there were several and long gaps in the time sequence data and hence it was difficult to derive the 
power spectra precisely. 
The studies related to intensity oscillations of the XBPs and their contribution to coronal 
heating were not well discussed with Yohkoh/SXT observations.  The X-ray Telescope on Hinode, XRT, has
made such long and continuous high temporal and spatial resolution time sequence observations of XBPs.
In addition the
angular resolution of XRT is 1", which is almost three times better than that of Yohkoh/SXT instrument.  Due to 
a wide coronal temperature coverage achieved with XRT observations, 
for the first time the XRT can
provide a complete dynamical evolution of the XBPs. The
study of the spatial and temporal relationship between the solar coronal XBPs and the photospheric and
chromospheric magnetic features is an important issue in physics of the Sun.
The Hinode/XRT observations provide an opportunity to investigate and understand
more deeply into the dynamical
evolution and nature of the XBP than has been possible to date and to determine their connection to the large-scale
magnetic features.  Such a high resolution observations and investigations would certainly be
helpful in understanding the role of oscillations and nature of the waves associated
with XBPs to heat the corona.
        
	In this paper, we report the analysis of 14 XBPs and 2 background coronal regions
showing different emission levels selected from 7-hour time sequence observations of soft X-ray images 
obtained on April 14, 2007.  We discuss the results of the periods of intensity oscillations associated 
with XBPs of different emission levels.

\section{Methods and results}


We use a 7-hour (from 17:00 UT to 24:00 UT) time sequence of soft X-ray images
obtained on April 14, 2007 from X-Ray Telescope (XRT) on-board the Hinode mission. Images
were obtained at time
cadence of 2 min, but with a short data gap around 18:00 UT.  The images have 
been observed through a single X-ray Ti\_poly filter in a quiet region near to the center of the solar
disk and the image size is 512"x512" with a spatial resolution of 1.032"/pixel.  Figure 1, shows a sample of an
X-ray image from the time sequence
illustrating all the XBPs, which we will be discussing in more detail below.
  We have identified and selected 14 XBPs and 2 background coronal regions of the images around 17:00 UT
for the analysis.  We have marked all the 14 XBPs on the image in Fig. 1 as
xbp1, xbp2,............, xbp14 and 2 background coronal regions as xbp15 and xbp16.

\begin{figure}
   \centering
  \includegraphics[width=13.0cm]{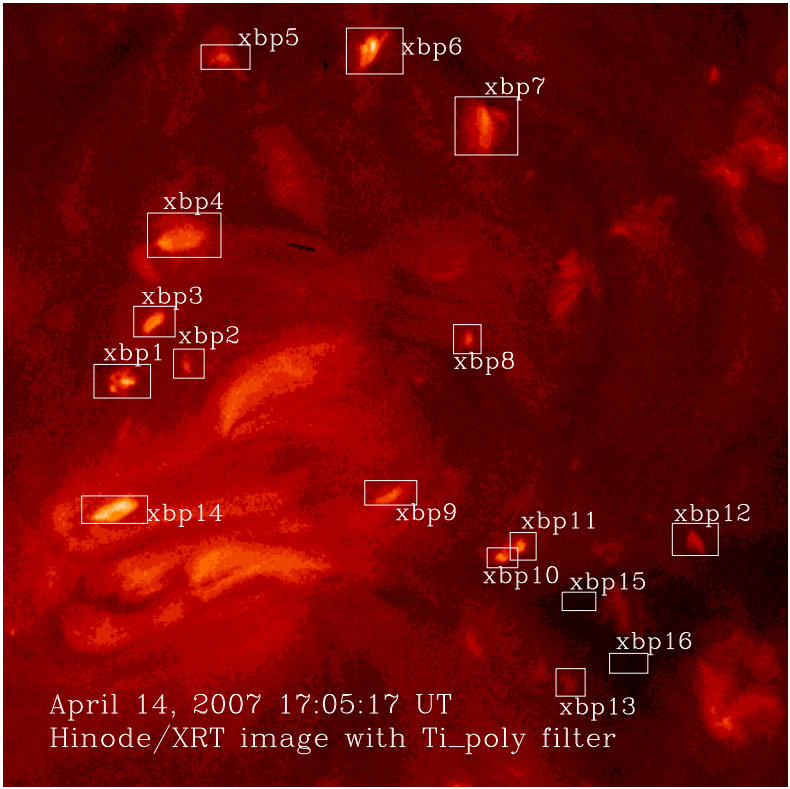}
   \caption{A sample of an image from the time series obtained by Hinode/XRT on April 14, 2007 at
17:05:17 UT.  Where xbp1, xbp2,...........xbp14 are the X-ray bright points and xbp15 and xbp16 are
the background coronal regions selected for the analysis.}
\label{FigVibStab}
    \end{figure}
We have used the routine xrt\_prep.pro in IDL under SSW tree to calibrate the images and this routine
does many corrections such as : (i) removal of
cosmic-ray hits and streaks (using the subroutine xrt\_clean\_ro.pro), (ii) calibration of read-out signals,
(iii) removal of CCD bias, (iv) calibration for dark current, and (v) normalization of each image for exposure time.
On the calibrated images we have put the rectangular boxes covering the selected XBPs and derived
the cumulative intensity values of the XBPs by adding up all the pixel intensity values.  The light curves of 
all the XBPs and background regions have been derived.  The light curves of the background 
coronal emission show that the fluctuations due to the background coronal emission is very small compared 
to the intensity variations at the sites of XBPs.
However, we assumed that the background coronal emission is more or less constant and
subtracted the intensity values of XBPs, point-by-point, from the background coronal intensity values
(xbp15 and xbp16) in all the images.

 \begin{figure}
   \centering
  \includegraphics[width=13.0cm,height=9.5cm]{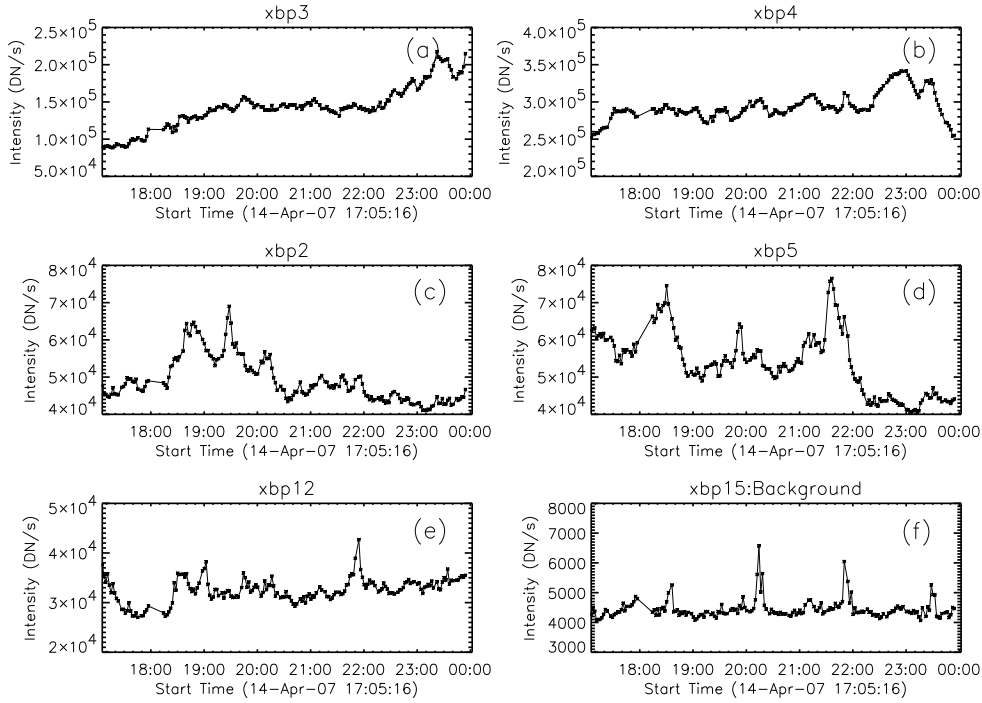}
   \caption{Time series of XBPs and background coronal region (xbp2, xbp3, xbp4, xbp5, xbp12 \&
xbp15 as shown in Fig.1) of April 14, 2007, where (a)
xbp3 \& (b) xbp4: Class I; (c) xbp2 \& (d) xbp5: Class II and (e) xbp12: Class III and (f) xbp15: background coronal region.}
\label{FigVibStab}
    \end{figure}

The light curves of 5 XBPs (xbp2, xbp3, xbp4, xbp5, \& xbp12) and one background
coronal region (xbp15) are shown in Fig.2 for comparitive study among the selected XBPs.
The intensity oscillation is seen in all the light curves of the XBPs.    
There is a short data gap in all 
the time series around 18:00 UT due to lack of observations.  The light curve of xbp3 is shown 
in Fig.2(a) which indicates a steady increase in brightness with intensity oscillation.  While 
Fig.2(b) shows the brightness variation of xbp4
and it is similar to xbp3 in the intensity variations.  Figure 2(c) illustrates
the light curve of xbp2
exhibiting oscillation with a strong intensity enhancement in the time period from 18:30 UT to 20:30 UT.
The light curve of xbp5 is shown in Fig.2(d) and shows the intensity oscillations similar to other XBPs 
but with a strong intensity enhancement 
at two locations between 18:00 - 19:00 UT and 21:00 - 22:00 UT.  
Similarly the light curve of xbp12 is presented in Fig.2(e) and the intensity oscillations are clearly seen.  
The light curve of the background coronal 
region, xbp15, is also shown in Fig.2(f) which does not show
much intensity fluctuations compared to XBPs, except for sharp intensity brightenings about 
every 80-100 mins.

  In order to determine the periods of intensity oscillations, we have done a power
spectrum analysis of all the XBPs and coronal background regions.
The
power spectra of 5 XBPs and one background coronal region are shown in Fig.3.  Figure 3(a) shows
the power spectrum of xbp3 with three significant peaks at the beginning from low frequency values
which correspond to the periods of 139 min, 37 min \& 24 min.  
Figure 3(b) illustrates the 
power spectrum taken for the time series of xbp4 indicating many significant peaks at 
various frequency values which correspond to the periods of
111 min, 37 min, 23 min, 17 min etc.
The power spectrum of xbp2 is shown in Fig.3(c) and represents two prominent peaks corresponding to the 
periods of 69 min
and 28 min.  The power spectrum of xbp5 is demonstrated in Fig.3(d) representing several
periods of oscillation, namely 133 min, 33 min and 13 min.  Similarly, in Fig.3(e) the power spectrum of xbp12
is plotted which shows a variety of periods of 133 min, 33 min, 17 min and 9 min. 
The power spectrum of one of the
background regions (xbp15) is also shown in Fig.3(f) and has several prominent peaks corresponding
 to the
periods of intensity oscillation as 81 min, 31 min, 19 min, and 8 min.  The power
spectra of all the XBPs reveal a wide variety of time scales right from a few minutes to hours and
some periods of oscillation are similar.
\begin{table}
\caption{Classification of XBPs}             
\label{table:1}      
\centering                          
\begin{tabular}{c c c c c}        
\hline\hline                 

XBPs Class & XBPs & XBPs mean brightness & XBPs brightness/Background brightness & Mean Ratio \\
           &      &   (DN/s)        &  (where background brightness=4440 DN/s) &  \\

\hline                        

   Class I  &  xbp3  & 157389  & 35.45 &  \\      
            &  xbp4  & 288013  & 64.87 &    \\
            &  xbp6  & 145266  & 32.72 &  42.33  \\
            &  xbp7  & 162340  & 36.56 &   \\
            &  xbp14  & 186745  & 42.06 &   \\
   Class II  & xbp2  & 53573  & 12.07 &   \\
             & xbp5  & 58236  & 13.12 &  13.39  \\
             & xbp1  & 65445  & 14.74 &    \\
             & xbp13  & 60567  & 13.64 &   \\
   Class III  & xbp12  & 36978  & 8.33 &   \\
              & xbp10  & 26789  & 6.04 &  7.39  \\
              & xbp8   & 34676  & 7.81 &   \\
              & xbp11  & 37873  & 8.53 &   \\

\hline                                   


\end{tabular}
\end{table}

 \begin{figure}
   \centering
  \includegraphics[width=11.0cm,height=10.5cm]{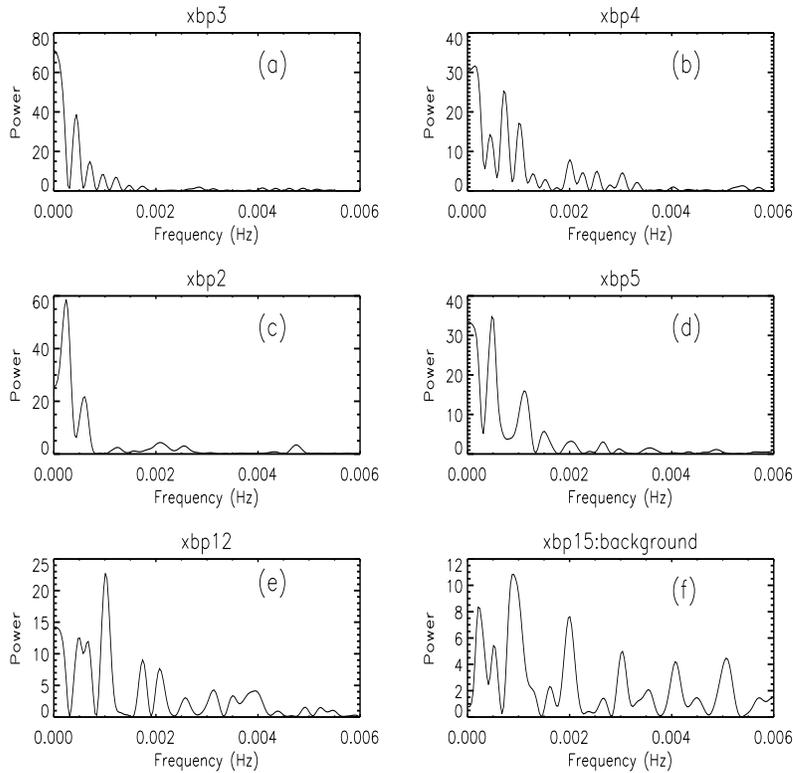}
\caption{Power spectra derived for the time series of XBPs and background coronal region (xbp2, xbp3,
xbp4, xbp5, xbp12 \& xbp15:background
as shown in Fig.2) of April 14, 2007, where (a)
xbp3 \& (b) xbp4: Class I; (c) xbp2 \& (d) xbp5: Class II and (e) xbp12: Class III and (f) xbp15: background coronal region.}
\label{FigVibStab}
    \end{figure}


\section{Discussions and conclusions}


The XRT data shows that XBPs tend to produce small and
large time scale fluctuations in their intensity and some periods of intensity oscillation are similar
in all the XBPs.  The periods observed with XRT data ranges from a few minutes to 
hours and these findings will have good agreement with the results of Strong et al. (1992) derived from the 
analysis of full-disk images obtained by Yohkoh/SXT experiment.

Although at first sight the light curves of 14 XBPs seem to be amazingly diverse in their pattern during
evolution, we find that they can broadly be grouped into three classes
depending on their emission level.  The Class I XBPs
show a very large intensity enhancement, whereas the Class II XBPs show moderate brightness enhancement
 and the Class III XBPs show only a marginal intensity enhancement during their dynamical evolution.
  We have calculated the mean intensity
of XBPs from their time series belong to Class I, Class II and Class III and the background coronal regions and
tabulated in Table 1.  We determined the mean intensity ratio between the different
classes of XBPs and the background coronal emission and the ratios have been listed in Table 1.  The mean
background intensity value of xbp15 and xbp16 derived over the time sequence is 4440 DN/s.  It is seen
from the light curves (Fig.2) and Table 1 that the XBPs will fall into 3 classes depending on the brightness
level during their dynamical evolution.  Class I XBPs are almost 42
times brighter than the background coronal emission, whereas the Class II and III XBPs are 13 and 7 times brighter
than background coronal emission respectively.  

We conclude that the analysis of long time series observational data of the XBPs is thus
promising and future work will tell more
about the dynamical nature and the physical properties of different oscillations and waves associated with the XBPs.
Since the periods of intensity oscillation in all the three cases of XBPs seem to be similar
and this can be taken as an evidence that heating mechanisms in all the three cases of XBPs are
similar.  XBPs exhibit a
wide variety of time scales ranges from a few minutes to hours in their intensity variations and the 
periods are almost similar in all the cases of XBPs and thus seems to be 
independent of the differences in the brightness enhancement.
The XBPs are the sites where intense brightness
enhancement is seen, and the brightness oscillates with different periods.
It suggests that the regions of intense vertical magnetic field strength coincide with regions
that are bright indicating non-radiative heating, irrespective of the sizes of these structures.  A comparison
study between the XBPs and underlying photospheric magnetic features has suggested that the 
horizontal component of the magnetic field may be playing an important role in driving the brightening of an 
XBP (Kotoku et al. 2007).  Therefore the possible reason
for the existence of different classes of XBPs with similar
periods among all the XBPs may be
related to the different strengths
of the magnetic field with which they have been associated.  


\begin{acknowledgements}
Hinode is a Japanese mission developed and launched by ISAS/JAXA,
collaborating with NAOJ as
a domestic partner, NASA and STFC (UK) as international partners. Scientific operation of the
Hinode mission is conducted by the Hinode science team organized at ISAS/JAXA. This team mainly
consists of scientists from institutes in the partner countries. Support for the post-launch
operation is provided by JAXA and NAOJ (Japan), STFC (U.K.), NASA (U.S.A.), ESA, and NSC (Norway).

 One of us (RK) wishes to thank Tetsuya Watanabe for discussion on several occasions on the topic
of XBPs and for his valuable information at the beginning in using the Hinode data.  Many thanks to Katharine
Reeves who clarified and helped in solving many problems to access to the XRT images and in their analysis. In
addition she had taken pain to read through the draft manuscript and given valuable suggestions.  RK wish to express
sincere thanks to Ed DeLuca for his valuable suggestions and discussion on the
XRT observations.  The authors are grateful to the unknown referee and to Hardi Peter (Editor, A \& A) 
for many useful comments, suggestions, and constructive remarks 
which helped considerably in improving the manuscript.
 
\end{acknowledgements}

\end{document}